\documentclass[RNAAS]{aastex62}
\usepackage{CJK}  
\usepackage{amsmath}
\graphicspath{{./}{figures/}}

\received{xx xx, 2019}
\revised{xx xx, xxxx}
\accepted{xx xx, xxxx}
\submitjournal{Research Notes}

\shorttitle{2I in K-band}
\shortauthors{Lee}

\begin{document}
\begin{CJK*}{UTF8}{bsmi}

\title{FLAMINGOS-2 infrared photometry of 2I/Borisov}

\correspondingauthor{Chien-Hsiu Lee}
\email{lee@noao.edu}

\author[0000-0003-1700-5740]{Chien-Hsiu Lee (李見修)}
\affil{NSF's National Optical Infrared Astronomy Research Laboratory, Tucson, AZ 85719, USA}

\author{Hsing-Wen Lin (林省文)}
\affil{University of Michigan, Ann Arbor, MI 48109, USA}

\author{Ying-Tung Chen (陳英同)}
\affil{Academic Sinica Institute of Astronomy and Astrophysics, Taipei, 10617, Taiwan}

\author{Sheng-Feng Yen (顏聖峰)}
\affil{Academic Sinica Institute of Astronomy and Astrophysics, Taipei, 10617, Taiwan}




\section{} \label{sec:intro}


2I/Borisov was first seen by Gennady Borisov by end of August 2019, and its extra-solar origin was established with more than a hundred astrometric observations over the next ten days \cite{2019NatAs.tmp..467G}. 2I/Borisov is the second interstellar object (ISO) after 'Oumuamua \citep{2017Natur.552..378M}, but differs from 'Oumuamua drastically with its extensive cometary activity. This immediately attracted spectroscopic follow-up observations, which result in the detection of CN emission \citep{2019ApJ...885L...9F}, [OI] as an indication of water \citep{2019arXiv191012785M}, and possibly C2 \citep{2019arXiv191003222K,2019A&A...631L...8O}. Besides the spectral lines, another key ingredient to understand the nature of this comet is its size. However, due to its cometary activity and extended coma in the optical, only rough estimates and upper limits can be made for 2I/Borisov, ranging in a wide spread from 0.7 to 3.8 km \citep{2019NatAs.tmp..467G,2019ApJ...885L...9F,2019ApJ...886L..29J,2019arXiv191014004B}.

\begin{figure*}[ht!]
  \hfill
  \includegraphics[width=0.95\textwidth]{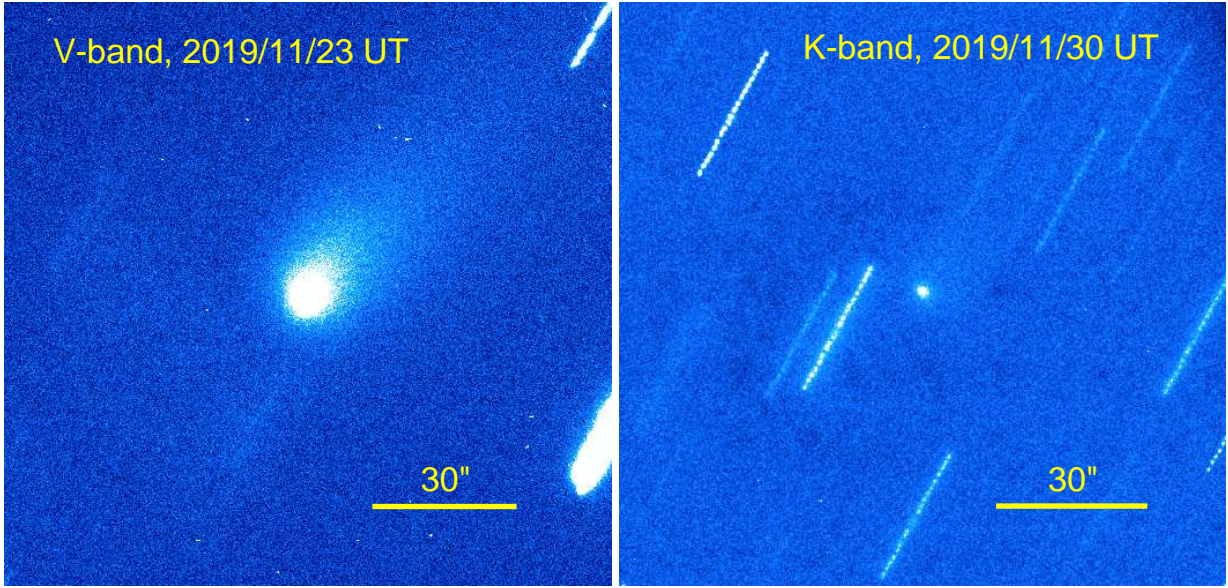}
\caption{Optical V-band (left) and infrared K-band (right) imaging of 2I/Borisov with North to the up and East to the left. The V-band imaging was taken with ALFOSC on-board the Nordic Optical Telescope on the night of November 23rd, 2019 UT, with 2$\times$ 300-second exposure (under NOT Fast-Track Service programme 59-419). The K-band imaging was taken with FLAMINGOS-2 on-board the Gemini South Telescope, with 20$\times$ 15-second exposure on the night of November 30th, 2019 UT (under Gemini Fast Turnaround program GS-2019B-FT-207). The optical imaging of 2I/Borisov significantly different from a point-source with the existence of an extended coma and tail, while the infrared imaging shows no apparent coma or tail and suggests a relative small nucleus.}
\label{fig.im}
\end{figure*}

It has been shown that observations at longer wavelengths (i.e. infrared) are less susceptible to the effect of coma, and can provide a better estimate of the size of the comet nucleus \citep[see, e.g.,][]{2013Icar..226.1138F,2017AJ....154...53B}. Here we present an estimate of the nucleus of 2I/Borisov from infrared observations. Using Gemini Fast Turnaround program (program ID: GS-2019B-FT-207), we observed 2I/Borisov with FLAMINGOS-2 on-board the Gemini South telescope, on the night of November 30th 2019 UT. We obtained 20$\times$ 15-second exposure in the K-band, which reveals a relatively small nucleus with no apparent coma or tail (Fig. \ref{fig.im}).

Using an aperture of 6.8 arcsec (equivalent to a projected radius of 10$^4$ km on the night of November 30th, 2019 UT), we found the comet had a K-band magnitude of 16.22$\pm$0.01 mag. Given the brightness, we can estimate the size of the comet using the procedure established by \cite{1991ASSL..167...19J}. First, we computed the absolute magnitudes H$_{K}$ using
\begin{equation}
H_K = m_K - 2.5 log_{10}(r_H \Delta) - 2.5 log_{10}(\Phi (\alpha)),
\end{equation}

where $r_H$ and $\Delta$ are the helio- and geo-centric distances (in a.u.), respectively and $\Phi(\alpha)$ is the phase function at phase angle $\alpha$. We assume a $\Phi(\alpha)$ in the form of $\beta \alpha$, with a generally used $\beta$ = 0.04 mag/degree. Given the absolute magnitude, we can estimate the size of the nucleus $r_e$ (in meters) by

\begin{equation}
r_n = \frac{1.496\times10^{11}}{\sqrt{p}} \times 10^{0.2(m_\odot-H_K)},
\end{equation}

where $p$ corresponds to the geometric albedo in the K-band and $m_\odot$ is the absolute magnitude of the Sun. There has not been much study of albedo in the infrared, so we assume $p$ = 0.07, a value determined by \cite{1992AJ....104.1611T} for comet 10P/Tempel 2. Using the above-mentioned equations, we estimate a comet nucleus size of $r_n$ = 1.5 km, comparable to but more stringent than the estimate from Keck AO imaging by \cite{2019arXiv191014004B}.


\acknowledgments

We are grateful to the staff at the Gemini Telescope, especially Dr. Morten Andersen and Dr. Hwihyun Kim, and staff at the Nordic Optical Telescope, especially Dr. Anlaug Amanda Djupvik, for supporting our observations.
Based on observations obtained at the Gemini Observatory and processed using the Gemini IRAF package, which is operated by the Association of Universities for Research in Astronomy, Inc., under a cooperative agreement with the NSF on behalf of the Gemini partnership: the National Science Foundation (United States), the National Research Council (Canada), CONICYT (Chile), Ministerio de Ciencia, Tecnolog\'{i}a e Innovaci\'{o}n Productiva (Argentina), and Minist\'{e}rio da Ci\^{e}ncia, Tecnologia e Inova\c{c}\~{a}o (Brazil).
Based on observations made with the Nordic Optical Telescope, operated by the Nordic Optical Telescope Scientific Association at the Observatorio del Roque de los Muchachos, La Palma, Spain, of the Instituto de Astrofisica de Canarias.

\end{CJK*}

\begin{thebibliography}{}

\bibitem[Bauer et al.(2017)]{2017AJ....154...53B} Bauer, J.~M., Grav, T., Fern{\'a}ndez, Y.~R., et al.\ 2017, \aj, 154, 53

  
\bibitem[Bolin et al.(2019)]{2019arXiv191014004B} Bolin, B.~T., Lisse, C.~M., Kasliwal, M.~M., et al.\ 2019, arXiv e-prints, arXiv:1910.14004

\bibitem[Fern{\'a}ndez et al.(2013)]{2013Icar..226.1138F} Fern{\'a}ndez, Y.~R., Kelley, M.~S., Lamy, P.~L., et al.\ 2013, \icarus, 226, 1138


\bibitem[Fitzsimmons et al.(2019)]{2019ApJ...885L...9F} Fitzsimmons, A., Hainaut, O., Meech, K.~J., et al.\ 2019, \apjl, 885, L9
     
\bibitem[Guzik et al.(2019)]{2019NatAs.tmp..467G} Guzik, P., Drahus, M., Rusek, K., et al.\ 2019, Nature Astronomy, 467

\bibitem[Jewitt(1991)]{1991ASSL..167...19J} Jewitt, D.\ 1991, IAU Colloq. 116: Comets in the Post-halley Era, 19

  
\bibitem[Jewitt, \& Luu(2019)]{2019ApJ...886L..29J} Jewitt, D., \& Luu, J.\ 2019, \apjl, 886, L29

  
\bibitem[Kareta et al.(2019)]{2019arXiv191003222K} Kareta, T., Andrews, J., Noonan, J.~W., et al.\ 2019, arXiv e-prints, arXiv:1910.03222

  
\bibitem[McKay et al.(2019)]{2019arXiv191012785M} McKay, A.~J., Cochran, A.~L., Dello Russo, N., et al.\ 2019, arXiv e-prints, arXiv:1910.12785

 \bibitem[Meech et al.(2017)]{2017Natur.552..378M} Meech, K.~J., Weryk, R., Micheli, M., et al.\ 2017, \nat, 552, 378

  \bibitem[Opitom et al.(2019)]{2019A&A...631L...8O} Opitom, C., Fitzsimmons, A., Jehin, E., et al.\ 2019, \aap, 631, L8

    \bibitem[Tokunaga et al.(1992)]{1992AJ....104.1611T} Tokunaga, A.~T., Hanner, M.~S., Golisch, W.~F., et al.\ 1992, \aj, 104, 1611

  
\end{thebibliography}
\end{document}